\documentstyle[amssymb,12pt,thmsa]{article}
\input{tcilatex}
\begin{document}

\begin{center}
{\bf Four-dimensional Matter}

\bigskip 

\strut {\sl A.M. Gevorkian , R. A. Gevorkian* }

{\sl Institute of Physical Research, 378410, Ashtarak-2, Republic of Armenia}

{\sl *Institute of Radiophisics and Electronics, 378410, Ashtarak-2,
Republic of Armenia}

\bigskip 

\smallskip 

{\bf Abstract}
\end{center}

{\it Four-dimensional mass is determined in four-dimensional
pseudo-Euclidean space as a physical invariant of that space. That invariant
is discussed as an invariant of electromagnetic type. Finally, equations of
Maxwell type are obtained for mass fields.}

\bigskip 

In Okun's publication ''What is mass?'' [1] the formula

\begin{equation}
m^2=\frac{U^2}{c^4}-\frac{P^2}{c^2}  \label{eq}
\end{equation}

and comments on it looked very interesting to us for two reasons:

1) as a four-dimensional matter (in [1] it is defined as ''simple mass'').

The analysis of non-stationar phenomena and processes lead V.A. Ambartsumian
to the idea of a possible new state of matter. In works [2], [3] and [4] the
proto-star matter is discussed as a possible presentation of this state.

There are also a few works by R. Muradyan ([5] and [6]) on unusual
manifestations (quantum) of macroscopic star systems (two-dimensional
hadron).

Publication [1] notes that during fragmentation four-dimensional mass is not
additive. The process of fragmentation continues to the point of occurrence
of elementary particles (electron, proton) with emission of energy. Later
the usual three-dimensional matter is constructed from those particles,
which is the process of three-dimensionalization. The simplest form of this
matter is the hydrogen atom. The further cooling of the environment by means
of condensation, leads to the formation of usual star matter and its
systems, and these systems may be unstable. In that case the famous
disagreement between Ambartsumian and his opponents regarding the concept of
formation of star systems becomes unimportant. We would like to remind that
the first member of expression (1) is a square of a scalar, and the second
member is a scalar - a module of square of a three-dimensional vector.

For compensation of the deficit of the mean cosmic density of matter some
scientists proposed assumptions such as the ''spontaneous birth of matter''
(Hoyle), ''existence of galaxies of neutrino matter'' (Novikov), or
existence of dark matter due to objects like the ''black holes''.
Ambartsumian expressed opinions regarding this in [3], [4] and during
seminars where one of the authors was present, that the law of energy
conservation in its present form is incomplete. With respect to that we
assume that the invariant (1) is a scalar of anti-symmetric tensor, which is
created on the basis of three-dimensional (polar and axial) vectors.

2) as Lorentz invariant of square form with anti-symmetric tensor of
four-dimensional matter [7].

\begin{equation}
M_{ik}=\left| 
\begin{array}{cccc}
0 & U_x & U_y & U_z \\ 
-U_x & 0 & -P_z & P_y \\ 
-U_y & P_z & 0 & -P_x \\ 
-U_z & -P_y & P_x & 0
\end{array}
\right|
\end{equation}

Elements of this tensor can be considered as components of electric mass
(polar three-dimensional vector) and magnetic mass (axial three-dimensional
vector) fields.

With the help of this tensor we can construct invariant values with respect
to Lorentz modifications. These invariants have the following form:

\begin{equation}
M_{ik}M^{ik}=I_{1,}\text{ }e^{iklm}M_{ik}M_{lm}=I_2\text{ }or\text{ }%
P^2-U^2=I_{1,}\text{ }UP=I_2.
\end{equation}

$I_1$ corresponds to expression (1) and is discussed as Lagrangian of the
physical system and satisfies Lagrange equation which transforms into a
system of equations of Maxwell type. The four-dimensional form of these
equations is:

\begin{equation}
\frac \partial {\partial x^i}M^{ik}=j^k,\text{ }\varepsilon ^{iklm}\frac
\partial {\partial x^k}M_{lm}=0
\end{equation}

and in three-dimensional form it is

\[
divP=0\text{ \qquad }rotU=\frac 1c\frac{\partial P}{\partial t} 
\]

\begin{equation}
divU=\rho \text{ \qquad }rotP=\frac 1c\frac{\partial U}{\partial t}+\frac 1cj
\end{equation}

where $\rho $ is the source (charge) of electric-mass static field which has
the measurement of gr/cm and represents mass per unit length. And finally
the wave equation in four-dimensional form is $\square A=0$ , where $A$ is
the vector-potential of the electromagnetic-mass field, and $\square =-\frac{%
\partial ^2}{\partial x_i\partial x_k}$.

The source of field can be a mass string, as well as a mass ring. The
movement of mass along a string or a ring (any locked configuration) induces
magnetic-mass field of particular topology (toric, flat, etc).

It might be very interesting to place this in electric or magnetic
environments, for instance inside an electric charge. A similar thing exists
in the work of Ciama [6] in which he gives up the advantage of general
covariance in favor of the vector theory of gravitation. The parameters of
such an object can be considered when a quantum (calibration) theory is
created similar to quantum electrodynamics or theory of electro-weak
interactions. It is not excluded that these ideas may be virtual and may not
have anything in common with real physics.

\smallskip 

\bigskip 

\begin{center}
Literature
\end{center}

1. L. B. Okun ''What is mass?'' Physics - Uspekhi, 43(12) 2000.

2. Ambartsumyan V. A. Reports of the Academy of Science of Arm. SSR., 16,
97, 1953.

3. Ambartsumyan V. A. ''Phenomena of Continuous Emission and Sources of Star
Energy'', Notices of the Byurakan Observatory, 13, 1954.

4. Ambartsumyan V. A. ''Stars of T-Taurus and UV-Whale Type and the
Phenomenon of continuous emission'', Theses for the Symposium on
Non-stationar Stars. Theses for the 9th Congress of the International
Astronomic Society, 1955, 5, M., 1955.

5. Muradyan R., Astrophysics, 11, 237, 1975.

6. Muradyan R., Astrophysics, 13, 63, 1975.

7. Landau L. D., Lifshits E. M., ''Field Theory'', 1988.

8. Ciama D. W Roy Astron. Soc Monthly Notices 113, 34, 1953.

\end{document}